\documentclass[aps,prl,twocolumn,letter]{revtex4}

\usepackage{graphicx}
\usepackage{bm}

\newcommand{\be}{\begin{equation}}
\newcommand{\ee}{\end{equation}}
\newcommand{\bwt}{\begin{widetext}}
\newcommand{\ewt}{\end{widetext}}
\newcommand{\bea}{\begin{eqnarray}}
\newcommand{\eea}{\end{eqnarray}}

\begin{document}
\title{Optimized Noise Filtration through Dynamical Decoupling}
\author{Hermann Uys$^{1,2}$, Michael J. Biercuk$^{1,3}$, John J. Bollinger$^{1}$\\
$^1$ National Institute of Standards and Technology, 325 Broadway, Boulder, CO, 80305\\
$^2$ Council for Scientific and Industrial Research, Brummeria, Pretoria, South Africa\\
$^3$ Georgia Institute of Technology, Atlanta, Georgia\\
}

\date{\today}

\begin{abstract}
One approach to maintaining phase coherence of qubits through dynamical decoupling consists of applying a sequence of Hahn spin-echo pulses.  Recent studies have shown that, in certain noise environments,  judicious choice of the delay times between these pulses can greatly improve the suppression of phase errors compared to traditional approaches.  
By enforcing a simple analytical condition, we obtain sets of dynamical decoupling sequences that are designed for optimized noise filtration and are spectrum-independent up to a single scaling factor set by the coherence time of the system.  We demonstrate the efficacy of these sequences in suppressing phase errors through measurements on a model qubit system, $^{9}$Be$^{+}$ ions in a Penning trap. Our combined theoretical and experimental studies show that in high-frequency-dominated noise environments this approach may suppress phase errors orders of magnitude more efficiently than comparable techniques can.    
\end{abstract}

\maketitle

Developing improved strategies for maintaining quantum coherence in the presence of environmental noise is central to the advancement of quantum control experiments and applications.  The field of quantum information processing  stands to gain significantly from such strategies.  One approach to maintaining coherence consists of applying simple pulses intermittently \cite{Viola1998,Vitali1999,Uhrig2007,Cywinski2008,Uhrig2008} with computational control operations.  Similarly, it is possible to design the computationally relevant control operations themselves with the same objective in mind \cite{Viola1999, Gordon2008, Viola2008}.  These techniques are  commonly referred to as dynamical decoupling. One widely used scheme entails applying a sequence of Hahn spin-echo-style $\pi$-pulses \cite{Hahn1950}, successive pulses being separated by free-precession delays.  The precise durations of these inter-pulse delays that would provide optimum error suppression is currently a topic of many research efforts \cite{Khodjasteh2005, Uhrig2007,Cywinski2008,Uhrig2008}. The Uhrig dynamical decoupling (UDD) sequence \cite{Uhrig2007,Uhrig2008} has shown especially promising improvements over traditional approaches such as Carr-Purcell-Meiboom-Gill (CPMG) multipulse spin-echo \cite{Vandersypen2004} in certain noise environments.

In this letter we devise and test the noise-suppression capabilities of dynamical decoupling pulse sequences tailored for optimized noise filtration (OFDD sequences).  To that end, we enforce a simple analytical condition that does not depend on detailed knowledge of the noise spectrum, yielding, up to a scaling factor in time, a set of optimized pulse sequences.  Each member of the set of sequences is optimized for a unique choice of sequence duration and consequently each member has different \textit{relative} pulse locations.  In that sense the concomitant dynamical decoupling may be thought of as being ``locally optimized'' \cite{Biercuk2008}.  This should be contrasted with most other spin-echo-style dynamical decoupling, e.g. UDD, in which analytical considerations \cite{Uhrig2007} naturally lead to fixed relative pulse locations for all sequence durations.

Our studies demonstrate that OFDD sequences suppress errors comparably to, or better than, sequences with fixed relative pulse locations in arbitrary noise environments, making orders of magnitude gains in noise environments with strong high-frequency components and sharp high-frequency cutoffs.
The sharpness of the cutoff that allows improvement over CPMG by UDD was studied in \cite{Uhrig2008}. In low-frequency-dominated noise environments OFDD performs similarly to CPMG which is an effective slow-noise filter. 
Finally, we demonstrate the efficacy of phase-error suppression by testing OFDD sequences using trapped ions as a model system, showing strong agreement between theory and experiment.

We consider the time evolution of a two-level system (qubit) under the influence of random classical noise that causes an instantaneous deviation in frequency, $\beta(t)$, from the unperturbed qubit frequency, $\Omega$.
The dynamics of the qubit are governed by the Hamiltonian:
$\hat H = \hbar/2\left[\Omega + \beta(t)\right]\hat\sigma_z,$
where $\hat\sigma_z$ is the Pauli spin operator parallel to the quantization axis.
Adopt the following conventions for a decoupling sequence of $n$ $\pi$-pulses: let $\tau$ be the total dynamical decoupling sequence duration, equal to the sum of the free-precession delays between pulses, plus the sum of all $\pi$-pulse durations.  If the center of the $j^\textnormal{th}$ $\pi$-pulse occurs at time $t=t_j$ then let $\delta_j=t_j/\tau$ and let each $\pi$-pulse have duration $\tau_\pi$.
Assuming the qubit spin-vector is initially aligned along the $y$-axis we use the expectation value of $\hat\sigma_y$ in the frame rotating at frequency $\Omega$
as a measure of coherence
$W(\tau) =|\overline{\langle\sigma_y(\tau)\rangle}|=e^{-\chi(\tau)}, $%
where $\chi(\tau)$ is the coherence integral,
$\chi(\tau)=\frac{2}{\pi}\int_0^\infty S(\omega)\frac{F_n(\omega\tau)}{\omega^2}d\omega$ \cite{Uhrig2007,Cywinski2008,Uhrig2008}.
The angle brackets in the former indicate the quantum mechanical expectation value and the over-line an ensemble average.  $S(\omega)$  is the power spectral density of $\beta(t)$, $\omega$ is the angular frequency, and $F_n(\omega\tau)$ \cite{Biercuk2008, Biercuk2009} a filter function dependent on the $n$ $\pi$-pulses locations
\begin{equation}
\footnotesize
F_n(\omega\tau)=\left|1+(-1)^{n+1}e^{i\omega\tau}+2\sum\limits_{j=1}^n(-1)^je^{i\omega\delta_j\tau}\cos{\left(\omega\tau_\pi/2\right)}\right|^2. \label{filterf}
\end{equation}
\begin{figure}
\center
\includegraphics[angle=0, scale=0.215]{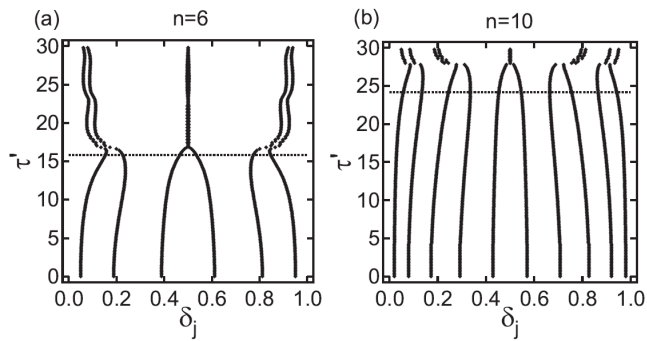}
\caption{OFDD sequences for (a) $n=6$ and (b) $n=10$ pulses. The horizontal axis
indicates the pulse locations, $\delta_j$, relative to the total sequence duration $\tau^\prime$ (vertical axis) that minimize $A_F(\tau^\prime)$.
The horizontal dashed line at (a) $\tau^\prime=15.8$ and (b) $\tau^\prime=24.2$ indicates when the filter function evaluated at the cutoff frequency is equal to one.
}\label{uloddpulses}
\end{figure}
\noindent Tailoring the filter function by choosing appropriate $\delta_j$ can lead to enhanced suppression of dephasing.

The coherence integral is smallest, and hence the error suppression strongest, when the overlap between the noise power spectrum and the filter function is minimized. Thus we enforce the condition that the area under the filter function $A_F(\tau)$ be minimized over a relevant frequency domain $[0,\omega_D]$ where $\omega_D$ is a potentially unknown high-frequency cutoff. This area depends both on the relative pulse locations, $\delta_{j}$, and the total sequence duration $\tau$; as such, optimization yields a \emph{set of pulse sequences}, each sequence optimized for a given value of $\tau$.  In the most general case $A_F(\tau)$ also depends on $\tau_{\pi}$, but the following discussion focuses primarily on the instantaneous pulse approximation ($\tau_\pi=0$), which is appropriate in qubit systems where $\pi$-pulse durations are small compared to the system's coherence time, e.g. \cite{Madsen2006,Press2008}.

Transforming to dimensionless units by defining $\omega^\prime\!=\!\omega/\omega_D$ and $\tau^\prime\!=\!\omega_D\tau$ we want
$
A_F(\tau^\prime)=\omega_D\int_0^{1}F_n(\omega^\prime\tau^\prime)d\omega^\prime\label{uloddcondition}
$
to be a minimum. No detailed knowledge of the noise spectrum enters into this condition and $\omega_{D}$ appears only as an overall scaling factor, so that the same set of sequences minimizes $A_F(\tau^\prime)$ for any $\omega_{D}$. In practice, therefore, the same set of sequences can be used for any noise environment
\emph{by scaling the duration of all sequences in the set by the same factor while maintaining the same relative pulse locations in each individual sequence}.

For fixed $\tau^\prime$ we determine the relative pulse locations that minimize $A_F(\tau^\prime)$  by running a numerical search routine. We use the UDD sequence as an initial guess at $\tau^\prime\!=\!0$ and the sequence obtained after convergence as the initial guess for the next choice of $\tau^\prime$, and so on.  In Fig.~\ref{uloddpulses}(a) and (b) we show the sets of OFDD sequences so obtained for a range of sequence durations and assuming, respectively, $n\!=\!6$ and $n\!=\!10$ instantaneous  $\pi$-pulses, symmetrically distributed around $0.5\tau^\prime$.

The observation that the condition of minimized $A_F(\tau^\prime)$ yields a set of pulse sequences which provide near-optimum performance in any noise environment constitutes the central result of this work.  These sequences lead to significant gains over existing ones in high-frequency-dominated noise environments as we show next. 

\begin{figure}
\center
\includegraphics[angle=0, scale=0.22]{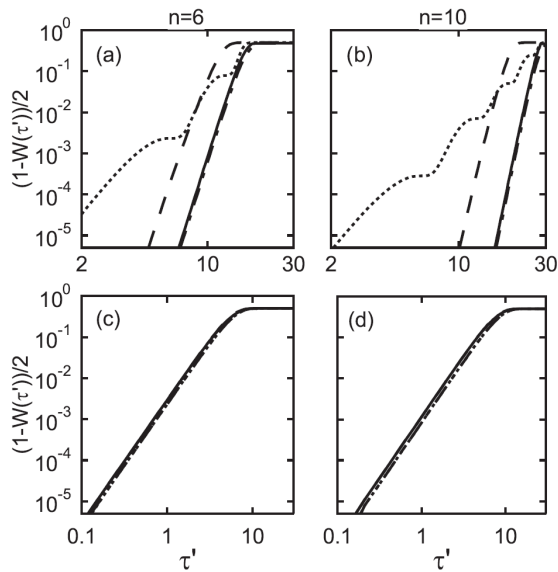}
\caption{Comparison of decoherence, $(1-W(\tau^\prime))/2$, due to phase errors when using $n=6$ (left) and $n=10$ (right)  pulse CPMG (dotted lines), UDD (dashed lines), OFDD (solid lines) and spectrum dependent LODD (dot-dashed lines) in an (a), (b) Ohmic and (c), (d) $1/f$ noise environment respectively. We assumed $\tau_\pi=0$ and used $\alpha\!=\!1$ throughout.}\label{Sims}
\end{figure}

We present simulations comparing the error suppression capabilities of OFDD sequences obtained by minimizing $A_F(\tau^\prime)$ only, to that of UDD and CPMG, as well as the sequences obtained by minimizing the spectrum dependent $\chi(\tau)$ directly (referred to as LODD - locally optimized dynamical decoupling \cite{Biercuk2008}).
In Figs.~\ref{Sims}(a)-(d) we plot simulations of the decoherence, $(1-W(\tau^\prime))/2$, having used $n\!=\!6$ and $n\!=\!10$ instantaneous $\pi$-pulses, and a dimensionless spectrum $S(\omega^\prime)/\omega_D=\alpha\omega^{^\prime\gamma}$, where $\alpha$ is the dimensionless noise strength.  We chose two limiting cases to illustrate the range of possible outcomes when using OFDD sequences. Those are an Ohmic, $\gamma=1$ (relevant to a spin-boson model for semiconductor quantum dots \cite{Uhrig2008}), and a ``$1/f$" $(\gamma=-1)$ spectrum respectively.  We assume an infinitely sharp cutoff for the Ohmic, but for the $1/f$ spectrum a soft high-frequency cutoff $\sim\omega^{\prime-2}$ and a sharp low-frequency cutoff at $\omega^\prime=0.001$ (to prevent spectrum divergence as $\omega^\prime\!\rightarrow\!0$).  
The main features of this comparison are that, in the high-frequency-dominated noise environments, Fig.~\ref{Sims}(a) and (b), the OFDD
approach achieves superior error suppression over the entire coherence time.  The OFDD error curve has roughly the same polynomial growth as that of
UDD, but extends the coherence time by a factor $\sim1.5$.  Our simulations show that these characteristics persist
independently of the noise strength and pulse number up to $\sim20$ $\pi$-pulses, beyond which the optimization becomes
numerically challenging.  Note that the OFDD result differs from the LODD approach by a relative error of only a few per cent (the two curves are nearly indistinguishable). In the
low-frequency-dominated $1/f$ environment, Fig.~\ref{Sims}(c) and (d), all four approaches perform similarly.  We have verified that similar improvements occur for supra-Ohmic noise spectra ($\gamma>1$) with sharp cutoffs, results are similar to CPMG for $\gamma<-1$, while intermediate benefits are achieved for white noise with a sharp cutoff.  As a rough measure of what constitutes a ``sharp" cutoff, we find empirically that significant benefits arise from our approach if the integrated noise power beyond a high-frequency peak is less than $\sim$10\% the total integrated noise power and contained within a frequency width one tenth of $\omega_D$.

These numerical results motivate us to develop an experimental procedure for implementing OFDD sequences.
To that end, consider the following analysis. The average relative delay between pulses is $\approx1/n$. At times $\tau^\prime\ll1$  all the phase factors in
Eq.~(\ref{filterf}) (within the domain $\omega^\prime \epsilon [0,1]$) are small compared to $\pi/2$, and the different contributions
in the sum add destructively (in particular $F_n(0)=0$).  
At times $\tau^\prime\gg1$ the phase factors in successive terms in the sum can differ by
order $\sim\pi$, and the terms in the sum may add constructively.  
The filter function then oscillates rapidly as a function of frequency around its average value $4n\!+\!2$ and the system is largely dephased.  The crossover between the two regimes occurs when $\omega_D\tau/n\approx\pi$, i.e., $\tau^\prime\approx n\pi$, leading to $F_n(\tau^\prime)\sim \mathcal{O}(1)$.  This is indicated by the horizontal lines in Figs.~\ref{uloddpulses}(a) and (b), designating $\tau^{\prime}=\tau_{F=1}^{\prime}$, where  $F_n(\tau^\prime) = 1$.

With these considerations in mind, appropriate scaling of OFDD sequences for a given experimental setting (i.e. an effective measurement of $\omega_D$) can be accomplished experimentally using a feedback routine as follows:\\
1) Numerically calculate the OFDD set for $n$-pulses in dimensionless units over a time domain $[0,2\tau^\prime_{F=1}]$.\\
2) Measure the coherence time $\tau_c$, i.e. the time for the error to increase to $1/e$ of its asymptotic value, for the $n$-pulse sequence, e.g. using CPMG or UDD.\\
3) Associate the pulse sequence (obtained in (1)) at $\tau^\prime_{F=1}$  with the measured $\tau_c$ as a first estimate of the scaling.\\
4) Find the sequence in the OFDD set that optimally suppresses errors for $\tau=\tau_{c}$ using a one-dimensional search algorithm  and experimental feedback.  Only a single sequence in the set should provide optimum error suppression when the $\tau_c$ is chosen as the sequence duration since the sequences are \textit{locally} optimized.\\
5) Scale the duration of all sequences in the set using the result of the above optimization.

In the above prescription, a single-parameter feedback algorithm, implemented at only one choice of sequence duration, simultaneously determines the scaling of all sequences within an analytically derived OFDD set. This stands in contrast to the $n$-dimensional feedback optimization routine (for $n$ pulses) used in our earlier work \cite{Biercuk2008} as a means of finding optimized sequences, and which must be repeated for each choice of sequence duration.

The generality of the OFDD approach discussed so far applies strictly only to the instantaneous $\pi$-pulse limit. When the pulse durations are nonzero and comparable to the inverse cutoff frequency, the relative time scale of
$\tau_\pi$-to-$1/\omega_D$ -- possibly an experimental unknown -- enters the problem. However, we find
empirically that (up to  $\omega_D\tau_\pi\approx1$) the pulse sequences obtained by accounting for $\tau_\pi$ in
minimizing $A_F(\tau^\prime)$ are nearly the same set of sequences as shown in Fig.~1, but contracted along the $\tau^\prime$ axis as compared to the instantaneous pulse case.  Hence the two scenarios differ primarily in the same scaling factor that the feedback is designed to determine, and the technique still works.  After scaling, significant deviation between the instantaneous and finite-duration cases occurs only at short times, when the total free-precession time is comparable to $n\tau_\pi$, but there the error suppression is extremely strong for any dynamical decoupling approach.

\indent We use $^9$Be$^+$ ions in a Penning trap as a model qubit system to demonstrate the proposed dynamical decoupling scheme.  We present only a brief summary of our experimental system, as it has been described in detail elsewhere \cite{Biercuk2008,Biercuk2009,Jensen2004}.  A few hundred to a few thousand ions are trapped and Doppler laser-cooled to $\sim\!1$ mK.
The $\sim\!124$ GHz (at 4.5 T), \small2s$\,^2$S$_{1/2} |m_I=3/2,m_J=-1/2\rangle=\left|\downarrow\right\rangle\longrightarrow|m_I=3/2,m_J=+1/2\rangle=\left|\uparrow\right\rangle$ \normalsize  spin-flip transition of $^9$Be$^+$ serves as a qubit.  Coherent rotations of the qubit are induced by a microwave field, producing a tunable Rabi flopping $\pi$-pulse duration, here 229 $\mu$s.
\begin{figure}
\center
\includegraphics[angle=0, scale=0.22]{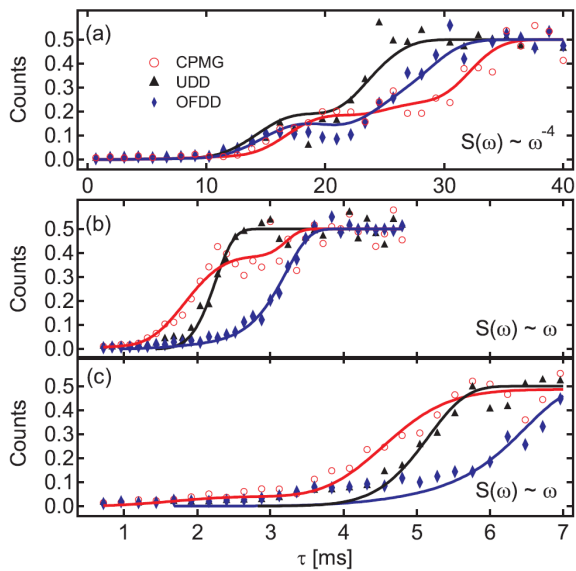}
\caption{Time evolution of the normalized fluorescence counts, equivalent to $(1-W(\tau^\prime))/2$, measured using $^9$Be$^+$ qubits in a Penning ion trap in (a) the ambient noise
environment and a synthesized Ohmic noise environment with (b) a frequency cutoff near $500$ Hz and (c) a frequency cutoff near $250$ Hz.
We used 6 $\pi$-pulses in each sequence and open circles represent CPMG, triangles UDD, and diamonds
OFDD.  The data are well described by the fits (solid lines) which use the analytical expression for the coherence, the measured noise spectra in both environments, and the noise strength as a free parameter.  We are at this time unsure of the cause of the minor discrepancy between the fits and the UDD and OFDD data in (c).  This may be the consequence of external noise influences not accounted for in the synthesized spectrum.
}\label{datafig}
\end{figure}
\indent Each experiment begins by optically pumping all ions into the state ($\left|\uparrow\right\rangle$) which is bright to cooling light fluorescence. The qubits are then rotated to lie along the $y$-axis by applying a $\pi/2$-pulse before initiating the decoupling sequence.  The decoupling sequence ends with a $\pi/2$-pulse that in the absence of dephasing rotates the qubits to the dark state ($\left|\downarrow\right\rangle$).  The accumulation of phase errors is manifested as nonzero fluorescence at the end of the experiment, due to a nonzero probability of qubit population in $\left|\uparrow\right\rangle$.  The normalized fluorescence count rate measured after application of a decoupling sequence is  a measure of this error and is given by
$(1-W(\tau))/2\label{erroreq}$. We achieve a minimum combined operational and measurement fidelity of $\sim99\;\%$ for sequences with $n\lesssim10$. A desired noise environment is synthesized by frequency-modulating the microwave drive as explained in \cite{Biercuk2009}. 

\indent Figure \ref{datafig} compares the experimental performance of different decoupling sequences measured in (a) the ambient noise environment of our trap, which approximately scales as $\sim1/\omega^4$ \cite{Biercuk2008,Biercuk2009},  (b) and (c) a synthesized Ohmic spectrum with a sharp cutoff around $500$ Hz and $250$ Hz respectively.  We used here the OFDD sequences shown in Fig.~\ref{uloddpulses}(a), having subdivided the interval $0<\tau^\prime<30$ into about 3000 intervals. After measuring a 6 $\pi$-pulse CPMG error curve we choose for each spectrum a fixed sequence duration $\tau\approx\tau_c$ ($\tau=20$ ms, $2.8$ ms and $5.5$ ms respectively) and perform a golden-section search with experimental feedback to determine which OFDD sequence of the set gives optimum error suppression for the selected $\tau$. The feedback algorithm converges to the optimized sequence within $\sim\!10$ iterations. (More specifically to a narrow band of sequences all of which  give an error indistinguishable within the measurement noise.  This band can in principle be made arbitrarily narrow with sufficient averaging during measurement.) This fixes the appropriate sequence associated with any other choice of $\tau$.  As anticipated from the simulations presented earlier, we obtain comparable results for the different decoupling techniques in the ambient noise environment with a soft high-frequency cutoff,
Fig.~\ref{datafig}(a). By contrast, in the Ohmic noise environment with a sharp cutoff around $500$ Hz, Fig.~\ref{datafig}(b), the OFDD sequence shows significant improvements over the other sequences even in the low-fidelity regime (decoherence $>1\;\%$).  To illustrate that the feedback algorithm finds the appropriate scaling for arbitrary cutoff we decrease the cutoff by a factor of two and repeat the experiment, Fig.~\ref{datafig}(c). This increases the coherence time of the system, but the benefit due to OFDD remains.  Note that despite having  finite-duration $\pi$-pulses $\omega_D\tau_\pi\approx0.7$ the technique still works for reasons explained earlier.

\indent In conclusion, we have presented a dynamical decoupling technique designed for optimized noise filtration by enforcing an analytical condition that does not rely on any detailed knowledge of the noise power spectrum.  The technique significantly reduces the burden of performing measurement-feedback-based optimization of decoupling sequences \cite{Biercuk2008}, as it relies on only a single parameter search at one point of the error curve. Under appropriate conditions it improves error suppression by several orders of magnitude compared to those of standard sequences.  Experimental measurements using a model quantum system and artificially engineered noise environments validate the predicted performance of these  sequences.

We thank W.M. Itano, N. Shiga and A.P. VanDevender for contributions to hardware and software developments that enabled this experiment, and Y. Colombe and D. Hume for helpful comments on the manuscript.  We acknowledge funding from IARPA and NIST. M.J.B. acknowledges support from IARPA and Georgia Tech., and H.U. acknowledges support from the CSIR.
This Letter is a contribution of NIST, not subject to U.S. copyright.

%\bibliography{spinecho}

\end{document}